# Work function and structures of (100), (110), and (111) gold surfaces at low and high oxygen-coverages

Yukio Watanabe[1,2*], S. Miyauchi[1], S. Kaku[1], T. Yamada[1], and A. Horiguchi[1]

[1]Kyushu University Open Design Lab, Shiobara, Minami-ku, Fukuoka, Japan 815-8540

[2]University of Hyogo Department of Electronics and Computer Science, Himeji, Hyogo 671-2201, Japan

The Work function ($\phi$) is fundamental for chemistry and electronics. Additionally, $\phi$ can be used to examine the validity of the theoretical surfaces by comparing it with experimental $\phi$, even in the absence of long-range orders. In the reported and present experiments, the difference in $\phi$ between pristine and oxygen-covered Au surfaces ($\Delta\phi$) is <1 eV at ≤1 ML (1 ML: one full-monolayer). Contrarily, the available density functional theory (DFT) reports $\Delta\phi \approx 3$ eV for Au(111) surfaces at 1 ML. Hence, we study structures of O-atom-covered Au(100), Au(110), and Au(111) surfaces using DFT. The calculated most stable structures show $\Delta\phi$ <1.1 eV at ≤1 ML and a nearly constant $\Delta\phi$ at > 1 ML, which match experiments and are confirmed using hybrid functional. These agreements result from the stability-criteria transition between low and high O-coverages, driven by the O-induced displacements of Au-atoms and the new surface structures at high O-coverages. The most stable structures exhibit molecule-like O arrangements at Au(111) surfaces at 1 ML and all surfaces at 2 ML; the former is considered chemisorption. At Au(111) surfaces, some structures containing O-atoms in subsurfaces have formation energies that approach those of the most stable structures, while the variation of these structures increases with surface size. Hence, mixing these structures with the most stable structures is believed to destroy long-range orders, which agrees with the experiments. The density of states at the surfaces calculated using the hybrid functional exhibit small bandgaps at the Au(100) and Au(110) surfaces at 1 ML.

## I. INTRODUCTION

Gold (Au), typically known for its chemical inertness, has shown potential for high catalytic activity when its surfaces are oxidized. This characteristic makes oxidized Au surfaces suitable for applications such as water splitting, offering a promising solution for the world's energy needs [1-3]. Furthermore, the crucial role of oxygen in chemical reactions at the Au surface has been well-documented, including its involvement in CO oxidation, alcohol oxidation, and propylene epoxidation [3]. Various methods, such as electrochemical processes, oxygen plasma, and ozone, have been successfully used to achieve the oxidation of Au surfaces [4-16]; the controversy over molecular adsorption vs. atomic chemisorption exists owing to the difference in temperature ($T$) and oxygen supply method: O ($O_3$) vs. $O_2$. The density functional theories (DFT) of oxygen-covered Au surfaces have clarified structural properties [17-21]. In particular, Sun et al. [17] and Daigle and BelBruno [21] clarified the dependence of the binding energy of oxygen atoms to Au surfaces on the occupation site at low oxygen coverage.

The work function ($\phi$) [22], which determines the electron ($e^-$) transfer between different materials, is fundamental for the electronic and chemical properties. Additionally, $\phi$ is a criterion for the validity of theoretical surface structures because of the sensitivity to surface-atom arrangements. Further, $\phi$ of Au serves as a standard reference for $\phi$ of other materials. However, reports on $\phi$ of oxygen-covered Au surfaces are scarce, except for the elaborate photoemission experiments by Saliba et al. [15] and Gottfried et al. [23]. Moreover, we found only one DFT study on $\phi$ [24].

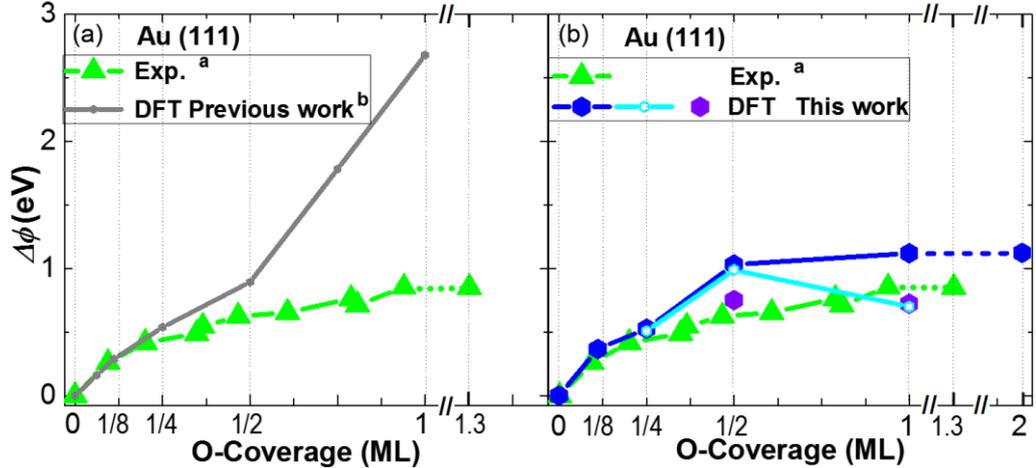

**Fig. 1.** Comparison of $\Delta\phi$ between experiments ([a] Saliba et al. [15]) and DFT: (a) previous DFT ([b] Stampfl and a coworker [24]) and (b) this work. Blue and light blue hexagons correspond to the structures that contain O-atoms only at surfaces and have the highest formation energy ($E^{Au-O}$). Violet hexagons are for the structures with O-atoms at surfaces and subsurfaces (details are in Sec. IIIC2). [a] Reproduced with permission from Surf. Sci., 410, 270 (1998) [15]. Copyright 1998 Elsevier. [b] Reproduced with permission from Phys. Rev. B. 76, 075327 (2007) [24]. Copyright 2007 American Physical Society.

Figure 1(a) shows the difference in $\phi$ between pristine Au and oxygen-covered Au surfaces ($\Delta\phi$) vs. O-coverage at Au(111) surface obtained by experiments [15] and DFT [24]. Here, 100% coverage means that the ratio of the number of O-atoms to the number of surface-Au-atoms is unity and corresponds to one O monolayer (1 ML). For Au(111) surface at ≤1.3 ML, $\Delta\phi < 1$ eV is experimentally found by Saliba et al. [15], while $\Delta\phi \approx 3$ eV at 1 ML is reported by the DFT of Stampfl and a coworker [24]. Additionally, the paper [24] reported that the binding energy between O and Au decreased steeply at >0.25 ML and became ≈ 2 eV at 1 ML. Because this energy is much lower than the binding energy per O-atom of $O_2$ ($E^{O-O}$), it is difficult to explain "chemisorption" in $O_2$ atmosphere reported by Huang et al. [5,6]. Here, the average binding energy of O-atoms to Au surface per O-atom ($E^{Au-O}$) has been defined as [17-21,24]

$$E^{Au-O} \equiv [E_{Au+O} - E^0_{Au}]/N_O - E_O,$$

where $E_{Au+O}$, $E^0_{Au}$, and $E_O$ are the energies of the O-covered Au surface, a pristine pure Au surface, and an isolated O-atom, and $N_O$ is the number of O-atoms in a supercell. At very low O-coverages, $E^{Au-O}$ is accurately the average *binding energy* of an O-atom to the Au surface per O-atom. However, at high O-coverages, $E^{Au-O}$ is the "average *formation energy* of O-covered surface per O-atom" because it includes the bonding energy between O-atoms.

$\Delta\phi \approx 3$ eV [24] has not been observed in other experiments either; Gottfried et al. [23] report $\Delta\phi \approx 1$ eV at ≤1.7 ML at Au(110) (i.e. Au(101)) surfaces. The present experiments also show $\Delta\phi < 1$ eV for the Au thin films, using ultra-high-vacuum (UHV) scanning force microscopy (SPM) [25]. Hence, the structures in ref. [24] may not correspond to the experimental ones.

This work resolves these issues; the present calculations will reveal new surface structures at high coverages that have $E^{Au-O}$ higher than $E^{O-O}$, which results from O-induced displacements of Au-atoms and molecule-like O-arrangements. Hence, O-chemisorption with high coverages is feasible even in $O_2$, as experimentally observed [5,6]. These calculations show $\Delta\phi \approx 1$ eV at 1 ML for all the Au(100), Au(111), and Au(110) surfaces (Fig. 1(b)), which agrees with the experiments of Saliba et al. [15], Gottfried et al. [23], and the present experiments.

Tran et al. [26] and De Waele et al. [27] show that the density functional theory (DFT) of $\phi$ of simple metals matches experimental values within 0.2–0.5 eV. However, for O-covered Au surfaces, $\Delta\phi$ ranges from 0 to 1 eV. Generally, a hybrid functional predict the electronic properties of oxides more accurately than DFT. Hence, we used DFT and a hybrid functional. The hybrid functional calculations of O-covered Au surfaces showed poor-metal-like electron density of states (DOS) at Fermi level ($E_F$), although the DFT showed metallic DOS at $E_F$.

## II. Methods
### A. Main points of methods
#### A1. Experimental

Polycrystalline Au films (100 nm thick) were sputter-deposited on $BaTiO_3$ single crystals (details are in Sec. IIB4). The lattice constant of the fcc surface of Au (4.07 Å [28])) is close to those of $BaTiO_3$ (4.04 Å and 3.99 Å [29]). The films were cleaned by burning out adsorbates with *atomic oxygen* (O) of 0.1 Pa at room temperature (RT) for three hours. The O-atoms were produced by electron cyclotron resonance (ECR) from cryogenically re-purified $O_2$ of initially 99.9999% purity (highest commercial grade) at a flow rate of 2 SCCM (cc/min). The validity of this procedure and the absence of impurities are explained in Sec. IIB4 [10,30,31].

When we assume the 0.1% conversion of $O_2$ to O, the cumulative area density of O-atoms that arrived at the Au surface in 3 hours was $6.5 \times 10^{17}$ cm$^{-2}$. If all these O-atoms remain at the surface, the coverage is 540 ML or 54000%. This estimate, the exposure by one order of magnitude longer than those in previous reports [15,23], and the reactivity of O-atoms much higher than $O_3$ molecules [15] suggest that the O-atom coverage was considered close to 100% (1 ML).

All the experiments were in the main UHV chamber of an SPM system (JEOL UHV-SPM 4610). $\phi$ was measured using Kelvin probe force microscopy (KPFM) *operated at a long distance* (5 nm) from Au surfaces in a true-noncontact mode based on evanescent Van der Waals force for macroscopic measurements. The probes of KPFM were cantilevers made of very heavy-doped *n*-type Si (resistivity of 10–25 mΩcm according to the manufacturer). More details are in Sec. IIB3.

#### A2. *Ab initio* calculations

*Ab initio* calculations used PAW potentials [32] implemented in VASP [33-35] with a plane wave energy cutoff of 650 eV. Because the accuracy of the lattice constant is fundamental for estimating oxygen positions, we used Perdew-Burke-Ernzerhof (PBE) functional optimized for solids (PBEsol) [36] as a generalized gradient approximation (GGA) DFT functional and the Heyd–Scuseria– Ernzerhof (HSE) functional for solids (HSEsol) [37] as a hybrid functional. According to Perdew et al. [36], PBEsol improves the accuracy of the equilibrium properties of solids and their surfaces.

We define the $c$-axis as perpendicular to the surface. The supercells for an O-covered Au system consist of 30 Å vacuum (Sec. IIB2), O-atoms, $m \times m \times l$ Au unit-cells that are $l$-layer Au slabs having $m \times m$ surface cells (Fig. 2); $9 \leq l \leq 41$, $1 \leq m \leq 4$, and the default is $l = 10$ and $m = 2$. Except for the cases of O-atoms in subsurface, appreciable changes in $\phi$ and $E^{Au\text{-}O}$ were not found in the calculations using supercells having larger surfaces.

We call these supercells "$m \times m \times l$ supercells" and call $m \times m$ surface cells "$m \times m$ supercells" in order to indicate that $m \times m \times l$ or $m \times m$ is the property of models and related with the initial structures before geometry optimization (relaxation of ion positions).

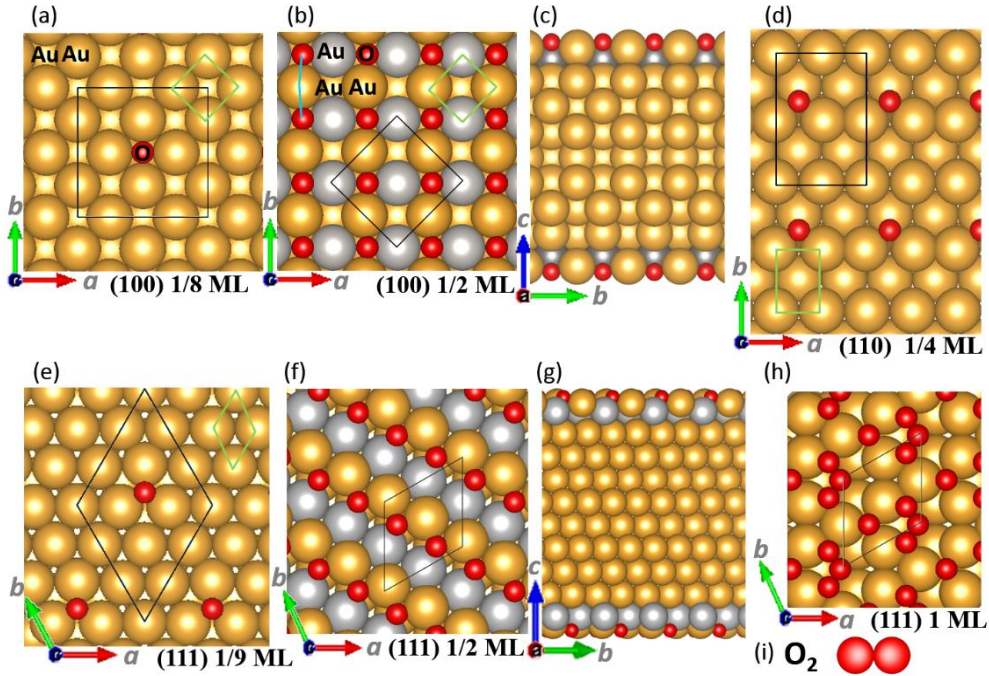

**Fig. 2.** Surfaces after geometry relaxation: top-views except for (c) and (f). (a)-(c) (100), (d) (110), (e)-(h) (111) ((h): with Au defects). (i) $O_2$. The O-coverage is indicated in each figure. In this paper, red and orange spheres show O- and Au-atoms, respectively, silver spheres show the Au-atoms at subsurfaces that existed at the top of pristine surfaces, and the diameter of O spheres is 1.8 Å, for which O-atoms in $O_2$ barely touch each other ((i)). In Figs. 2, 5-7, 9(a)-9(c), and 11, the diameter of Au spheres is 3.7 Å, for which inner Au-atoms barely touch each other (e.g., (e)). Black and green lines show supercells and (1×1) unit-cells of Au surface, respectively. Light blue lines show chain-like arrangements [17].

These supercells modeled O- covered Au(111), Au(110), and Au(100) surfaces, which are most intensively studied. $\phi$ of these surfaces was also hoped to approximately correspond to $\phi$ of thin films because a mixture of (100) and (111) orientations was reported for Au films that were sputter-deposited near 500 ºC [38].

Although partial relaxation allows thin supercells and is often used [24], it occasionally yields non-rigorous results. Hence, we relaxed all the ion positions without restriction except for the lattice constants of the slabs (supercells) by using thick supercells symmetric along the $c$-axis. After relaxation, all the calculated forces were lower than 1 meV/Å. In particular, we included the initial structures in which O-atoms were reported to be unstable or metastable [17,21], to enforce the extensive searches for stable structures.

The Monkhorst-Pack $k$-mesh for Brillouin-zone integration [39] in the geometry relaxation was denser than 24 points per $2\pi$ Å$^{-1}$. The validity of this mesh is reported [40-42]. For example, the mesh for Au(111) surfaces was $5 \times 5 \times 1$ for a $2 \times 2 \times 10$ supercell containing 40 Au-atoms (5.76 Å $\times$ 5.76 Å $\times$ 50 Å) and $3 \times 3 \times 1$ for a $3 \times 3 \times 10$ supercell (8.64 Å $\times$ 8.64 Å $\times$ 50 Å) containing 90 Au-atoms. A graphics-processing-unit acceleration [43,44] was employed. Ion positions are displayed using VESTA [45]. More details are in Secs. IIB1-IIB2.

$\phi$ is the energy required to move an $e^-$ from $E_F$ to the vacuum level [22] and is associated with dipole-like charge distribution at the surfaces. Figure 3 shows the $e^-$ density and the procedure to obtain $\phi$.

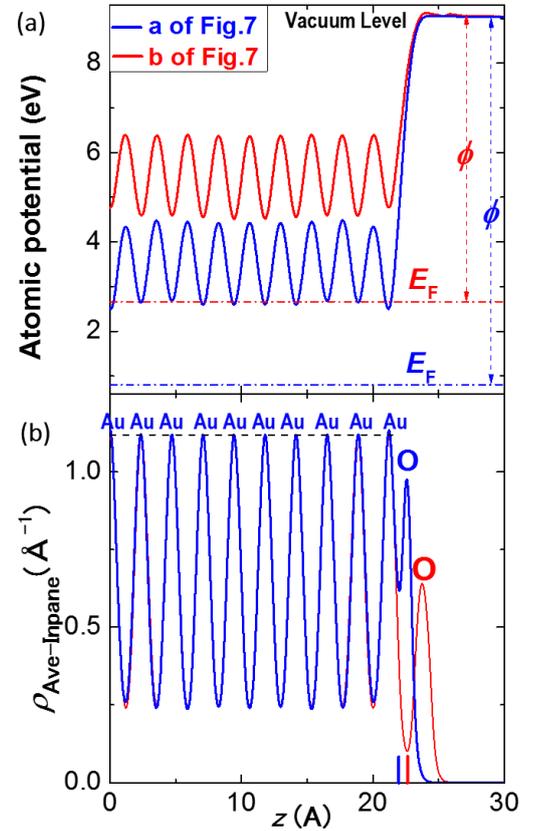

**Fig. 3.** Electrostatic potential (a) and plane-averaged $e^-$ density (b) of Au(111) surface at 1 ML O-coverage. Blue and red lines correspond to a high-symmetry structure and the most stable structure. Letters "Au" and "O" show the location of these ions. The $e^-$ occupation at O-atoms was obtained by integrating $e^-$ density beyond vertical lines at the abscissa of (b).

Unless otherwise stated, the PAW potentials for GGA used in this study are the default ones recommended by the VASP manual for solid metal oxides: Au PAW potential for Au and O_s PAW potential for O with a cutoff energy of 282.85 eV. Because adsorbed O-atoms can be different from O in solid oxides or have the properties of O in molecules, we also used the standard PAW potential for O ($O_{ST}$) with a cutoff energy of 400 eV. The overestimation of bond length by most GGA functionals is explicated by Perdew et al. [46]; hence, we used PBEsol GGA functional [36].

## A3. Performances of *ab initio* calculations and PAW

The deviation of the calculated lattice constant of bulk Au from the experimental value (4.070 Å) [28] was 0.010 Å (PBEsol), 0.020 Å (HSEsol), and 0.105 Å [24] (PW91), whereas PW91 GGA functional [47] is commonly used for O-covered Au surfaces.

$\phi$ of Au(100) surfaces calculated using PBEsol was insensitive to the number of layers $\geq 9$; $\phi$ was 5.189 eV, 5.159 eV, and 5.206 eV for nine-, ten-, and forty-layer supercells, respectively.

In comparison with O_s, $O_{ST}$ decreased the O-O bond length and increased the formation energy. In the case of $Au_2O_3$ ($Au_{16}O_{24}$), the $a$, $b$, and $c$ lattice constants calculated using PBEsol with $O_{ST}$ were 0.08%, 0.20%, and 0.18% shorter than with O_s, respectively, and the free energy calculated using PBEsol with $O_{ST}$ was by 0.32% greater than with O_s. The minimum Au-O distances using PBEsol with O_s and $O_{ST}$ were 1.962 Å and 1.959 Å, respectively. For an $AuO_2$ molecule, the bond length between O and Au ions and binding energy calculated using PBEsol with O_s and $O_{ST}$ were 1.769 Å, 3.68 eV, 1.767 Å, and 3.70 eV, respectively. The stable atomic configuration was a linear chain. These results on $AuO_2$ were consistent with those on Au-O molecules by Sun et al. [17].

For an $O_2$ molecule, the bond length and energies per O-atom calculated using PBEsol with O_s and $O_{ST}$ were 1.285 Å, 3.070 eV, 1.227 Å, and 3.323 eV, respectively. The bond length and energies per O-atom calculated using PW91 with O_s and $O_{ST}$ were 1.292 Å, 2.927 eV, 1.233 Å, and 3.158 eV, respectively. The experimental bond length and energy are 1.208 Å and 2.56 eV [48], with which PBEsol with $O_{ST}$ and PW91 with O_s agreed best, respectively. The overestimation of binding energies by the GGA functional in general is well known, and Perdew et al. reported 3.12 eV [46].

## B. Supplements to methods

**B1. Ion positions and parameters for HSEsol**

The lattice constants of a bulk Au obtained using PBEsol (4.080 Å) and HSEsol (4.090 Å) were similar. More importantly, $\phi$ for ion positions optimized with HSEsol was close to $\phi$ for those optimized with PBEsol; e.g., the HSEsol calculations using the ion positions optimized with PBEsol and HSEsol showed that $\phi$ was 5.177 eV and 5.157 eV for Au(100) surfaces, respectively, and 5.064 eV and 5.056 eV for Au(110) surfaces, respectively. Hence, the HSEsol calculations of $\phi$ of O-covered surfaces used the ion positions calculated using PBEsol.

The screening parameter ($\mu$) in HSEsol was the default value (0.2 Å$^{-1}$). We used the $k$-grid reduction option of the Hartree-Fock kernel (Nkred) of HSEsol. In the benchmark tests, Nkred $\leq 2$ did not compromise the accuracy of the calculation, and we used mainly Nkred = 2 in the in-plane directions.

**B2. Smearing, $e^-$ number convergence, and vacuum width for *ab initio***

The Gaussian functions with a temperature broadening ($\sigma$) of 0.05 eV were employed. The calculated values ($\phi$'s) using PBEsol ($\sigma$ = 0.05 eV) with a Gaussian and a Methfessel-Paxton functional were almost the same. For example, the Gaussian and Methfessel-Paxton methods resulted in $\phi$ = 5.634 eV and 5.634 eV for Au(100) surfaces at 0.25 ML O-coverage and $\phi$ = 6.305 eV and 6.298 eV for Au(111) surfaces at 0.5 ML O-coverage, respectively.

$\phi$ was sensitive to the $e^-$ distribution. Hence, in PBEsol calculations, the $e^-$-number convergence and the maximum plane-averaged $e^-$ density ($\rho_{\text{Ave-Inplane}}$ (Fig. 3)) in the substantial part of the vacuum were <10$^{-6}$ and <10$^{-8}$ Å$^{-1}$, respectively.

The vacuum width ($W_{\text{vac}}$) in the slabs is conventionally 15 Å [17,20,24] or, in careful studies, 20 Å [20]. However, the width of the evanescent region where $\rho_{\text{Ave-Inplane}}$ decreased to <10$^{-8}$ Å$^{-1}$ was approximately 6 Å at each side of the vacuum (in the case of 1 ML O on (111)), and its total was 12 Å. Hence, the standard choice of 15 Å seemed too short. In the tests of vacuum widths 20 Å and 30 Å, we observed mostly only a slight difference between these results, but we used 30 Å to secure accuracy. For example, in the following, the numbers without and with parentheses are $\phi$ (eV) at (111) surfaces for $W_{\text{vac}}$ = 30 Å and 20 Å using PBEsol, respectively (unless otherwise written, O_s PAW was used):

6.44 (6.45) for 1 ML O on $2 \times 2$ surface, 6.39 (6.39) for 1 ML O on $4 \times 4$ surface, 5.97 (5.97) for 1 ML O on $2 \times 2$ surface using $O_{ST}$, and 6.39 (6.46) for 2 ML O on $2 \times 2$ surface. The difference found in the last case was due to the structure changes, although in all the cases we optimized ion positions for $W_{vac}$ = 30 Å using the ion positions for $W_{vac}$ = 20 Å.

**B3. KPFM**

KPFM was in a long-distance mode with a low topological resolution. Further, we used the potential of the surface where the potential was uniform at a μm-scale in the middle of an Au film. Hence, KPFM worked as a conventional Kelvin probe.

The base pressure in the treatment chamber and the pressure in the main chamber were $2 \times 10^{-7}$ Pa and $<2 \times 10^{-8}$ Pa, respectively. As a test of the cleanness of the main chamber, clear images of individual atoms on a Si [111] surface measured by noncontact SPM showed no change for a month.

Cantilevers were from MikroMasch®. Removing oxides at the nano-scale pyramid at the top of the cantilevers is adequate for Kelvin probe measurements. For this, 150 ºC suffices, although it is far below the temperature for removing native oxides of a *bulk* Si ($\approx$1000 ºC). Hence, holding at 150 ºC in the UHV treatment chamber for six hours, we removed weak native oxides on the cantilever surface formed during the preservation in a dry box. This procedure is a standard one of the manufacturer (JEOL), and the validity was confirmed by measuring the correct contact potential of cleaned Si surfaces.

**B4. Thin film deposition and surface cleaning procedures**

Au films (100 nm thick) were sputter-deposited on $BaTiO_3$ single crystals at RT using 99.99% purity Ar gas and 99.9% purity Au target in a turbo-molecular-pumped vacuum chamber at a background pressure of $<10^{-3}$ Pa. Crystallographic orientations of the deposited Au films were undetected by X-ray diffractometry.

These films were taken out to air and then transferred to the treatment chamber of a UHV-SPM system. From the emission spectra, we determined that the emissions were of atomic O, except for negligible intensities of excited $O_2$ and $O_3$.

The elimination of adsorbates and the suppression of defect formation by this cleaning procedure were confirmed by X-ray photoemission [10,30]. The absence of interface material after this procedure was proven by the clear superconducting gap of $YBa_2Cu_3O_7$ [30]. Further, the ideal Schottky properties of the $SrTiO_3$:Nb surface obtained after this cleaning also verified the absence of defects and adsorbates [31]. The sufficient thermalization of the O gas was visible in the damage-free atomic steps of the substrate $BaTiO_3$ (Figs. 4(a) and 4(b)).

## III. RESULTS AND DISCUSSION
### A. Experiments of $\phi$

$\phi$ of polycrystalline Au thin films on $BaTiO_3$ single crystals was measured. $\phi$ of the KPFM probe that was a very heavily doped clean $n$-type Si ($\phi_{Si}$) corresponds to the energy level of the conduction band bottom of Si, which is 4.05–4.15 eV [22,49,50]. Hence, we use an average $\phi_{Si}$ = 4.10 eV. $\phi = \phi_{Si}$ − $\phi_{KPFM}$, where $\phi_{KPFM}$ is a surface potential given by KPFM.

As shown in Fig. 4(c), the sample was subjected to a heating cycle up to 100 °C for 10 min, followed by two heating cycles up to 140 °C with a 1-hour duration of >100 °C. In Fig. 4(c), $\phi$ is almost unchanged after the short 100 °C heating. $\phi$ decreased markedly after the $1^{st}$ 140 °C heating but only slightly changed by the $2^{nd}$ 140 °C heating cycle. The exponential fitting to these data indicates that $\phi$ becomes 4.95 eV after many heating cycles.

The reported experimental $\phi$ of polycrystalline Au thin films is 4.83 eV measured using a Kelvin method [51] and 5.1 ± 0.1 eV measured using a photoemission method [52]. The average of the reported $\phi$ of Au thin films is 4.965 eV.

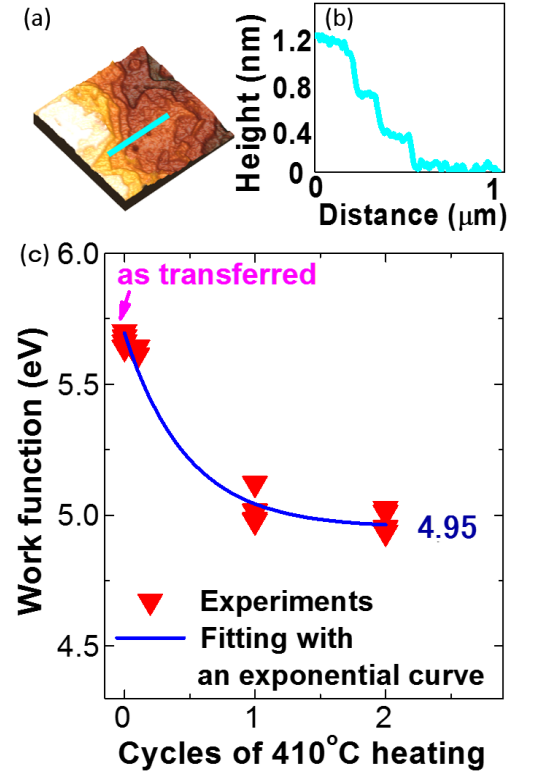

**Fig. 4.** (a) Surface topography and (b) cross-section analysis of substrate exposed to atomic O cleaning. We show the substrate because the Au surface was featureless. (c) Change in $\phi$ of O-atom-irradiated Au surface with annealing in UHV.

The experiments by Krozer and Rodah [53] and Ono and Cuenya [54] suggest that 150 $^\circ$C in UHV decomposes Au oxides at surfaces to yield Au. Further, $\phi = 4.95$ eV matches the average of the reported $\phi$ of Au thin films (4.965 eV) [51,52]. Hence, $\phi = 4.95$ eV is considered as $\phi$ of the Au film.

Section IIA1 showed that the O-coverage at the initial surface was approximately 1 ML. In this case, $\Delta\phi = 0.75$ eV between the initial and final surface in the experiments of Fig. 4(c) is $\Delta\phi$ between a 1-ML O-covered Au and a pure Au surfaces. This $\Delta\phi$ is consistent with $\Delta\phi$ obtained by the experiments of Saliba et al. [15] and Gottfried et al. [23].

## B. Pristine Au surfaces (*ab initio*)

In the calculations of defect-free pristine (pure) Au surfaces using $m \times m$ supercells ($m \leq 3$), surface reconstructions of Au surfaces were not observed when defects or O-atoms were absent.

The surface formation energy ($E_S$) of defect-free Au(100), Au(110), and Au(111) surfaces was 1.1 eV, 0.8 eV, and 0.3 eV, respectively; $E_S \equiv E_{Slab} - N_{Au} \times E_{Au}$, where $E_{Slab}$ and $E_{Au}$ are the energies of the Au slab and an isolated Au-atom, and $N_{Au}$ is the number of Au-atoms in a supercell. These values indicate that Au(111) surfaces were most stable, agreeing with the experiment on a freestanding Au film [55]. The surface formation energy slightly decreased as the size of the supercells increased; for the Au(100) surface, it was 1.1 eV for the supercells with 9 or 10 Au layers and 0.7 eV for the supercells with 40 Au layers.

| $\phi$ (eV) | HSEsol | PBEsol | PBE[a] | PBE[b] | PBE[c] | HSE[c] | Exp.[d] | Exp.[e] | Exp.[f] |
|---|---|---|---|---|---|---|---|---|---|
| (100) | 5.16 | 5.19 | 5.10 | 5.07 | | | | | 5.22 |
| (110) | 5.06 | 5.12 | 5.04 | 4.91 | | | | | 5.20 |
| (111) | 5.23 | 5.27 | 5.15 | 5.11 | 5.19 | 5.13 | | | 5.26 |
| Poly | | | | | | | 4.83 | 5.1±0.1 | |

**Table 1.** Comparison of *ab initio* calculated and experimental $\phi$ of defect-free Au surfaces. "Poly" means polycrystalline. [a] ref. [58]. [b] ref. [57]. [c] ref. [56]. [d] ref. [51]. [e] ref. [52]. [f] ref. [59].

In Table 1, $\phi$ calculated using PBEsol and HSEsol is consistent with reported *ab initio* and experimental $\phi$ [51,52,56-59]. Many papers and books list the experimental $\phi$ of Au single crystals. However, these data were citations from Hansson and Flodstrom [59], and we cited only ref. [59]. Concerning defects, we found that both $\phi$ was almost unchanged when one out of four surface Au-atoms were missing.

All the calculations and experiments of defect-free Au surface in Table 1 show that $\phi$ of Au(111) surface is the highest and $\phi$ of Au(110) surface is the lowest. This result corresponds to the surface Au-atom density, which is natural, considering the origin of $\phi$. Similarly, the difference in $\phi$ between PBEsol and PBE [56-58] in Table 1 can be attributed to the surface Au-atom densities because PBE overestimates lattice constants.

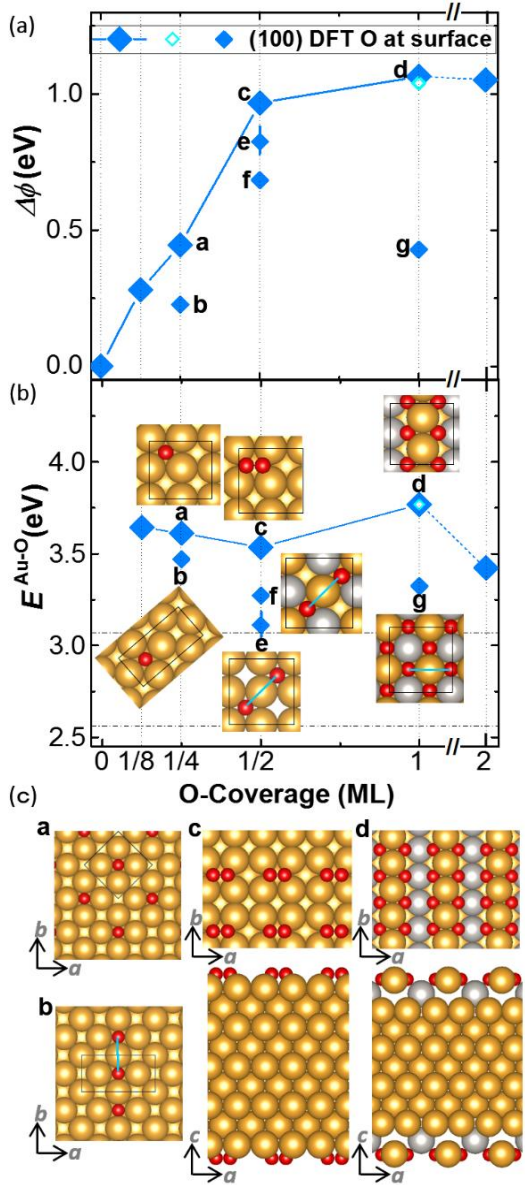

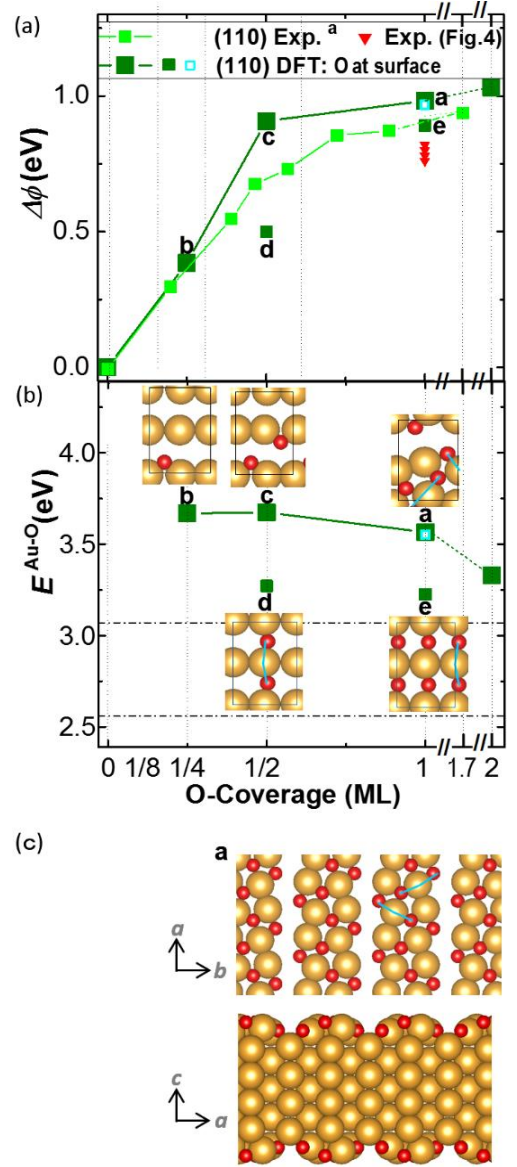

**Fig. 5.** (a) $\phi$ and (b) $E^{\text{Au-O}}$ at (100) surfaces. In (a) – (b), solid and open symbols represent the results using the O_s and O_{ST} potentials. "a" – "g" show the correspondence between data and surface structures. (c) Replots of the structures "a" – "d". The structures of "f" are also in Figs. 2(b) and 2(c).

**Fig. 6.** $\phi$ and $E^{\text{Au-O}}$ at (110) surfaces, plotted in the same manner as Fig. 5. In Figs. 5 and 6, the upper and lower horizontal black lines represent the experimental and theoretical $E^{\text{O-O}}$, and the light blue lines in pictures show chain structures [17]. [a] Reproduced with permission from Surf. Sci., 525, 197 (2003) [23]. Copyright 2003 Elsevier.

## C. O-covered Au surfaces (*ab initio*)

### C1. Au(100) and Au(110)

In Figs. 5(a) and 6(a), DFT of Au(100) and Au(110) surfaces exhibited $\Delta\phi$ <1.1 eV for ≤2 ML. In Figs. 5 and 6, outsized symbols correspond to the most stable surface (having the highest $E^{Au\text{-}O}$) at a given coverage, and $\Delta\phi$ of the most stable Au(110) surfaces by DFT agreed with experiments [23]. $\Delta\phi$ of thin films in Fig. 6 is close to $\Delta\phi$ reported by Gottfried et al. [23]. $E^{Au\text{-}O}$'s in Figs. 5(b) and 6(b) are consistent with those reported by Sun et al. [17] and Daigle and BelBruno [21].

The pictures in Figs. 5 and 6 show the surface structures of each data point. In these 2×2 supercell calculations, the most stable Au(100) surfaces show 2×2 superlattices at 0.5 ML and 2×1 superlattices at 1 ML. The most stable Au(110) surfaces show 2×2 superlattices at 0.5 ML and 1 ML.

The symmetries of the surfaces are better visible in expanded plots in Figs. 5(c) and 6(c). At 1-ML, the Au-atoms on the most stable surfaces deviate from their positions on pristine surfaces; at Au(100) surfaces, half of the Au-atoms reside at subsurfaces, and at Au(110) surfaces, the in-plane positions of Au-atoms markedly deviate from those of a pristine surface.

### C2. Au(111)

Because the experimental [15] and theoretical $\phi$ [24] significantly disagree at Au(111) surfaces (Fig. 1(a)), we examined the O-covered Au(111) surfaces extensively (Figs. 7-9). $\Delta\phi$ of the most stable Au(111) surfaces by DFT (PBEsol) agreed with experiments [15] (Fig. 1(b)).

The results for $\Delta\phi$ and $E^{Au\text{-}O}$ at 1 ML and 0.5 ML in Fig. 7 are mainly of 2×2 supercells. Even when calculated using bigger supercells, both $\Delta\phi$ and $E^{Au\text{-}O}$ were almost unchanged, and we observed that Au defects slightly increased $\Delta\phi$ and $E^{Au\text{-}O}$ (Fig. 7(a)).

For Au(111) surfaces, numerous structures that contained a part of the O-atoms at subsurfaces were found and showed low $\Delta\phi$ (purple symbols in Figs. 1(b) and 7). In Fig. 7(a), initial supercells are of five types: no O-atoms at subsurfaces, 1/4 of O-atoms at subsurfaces, 1/2 of O-atoms at subsurfaces, 3/4 of O-atoms at subsurfaces, and all O-atoms at subsurfaces. Further, each of these five types held different initial in-plane O-atom positions. After relaxing all the ion positions in these initial supercells without restriction, we obtained Fig. 7(a).

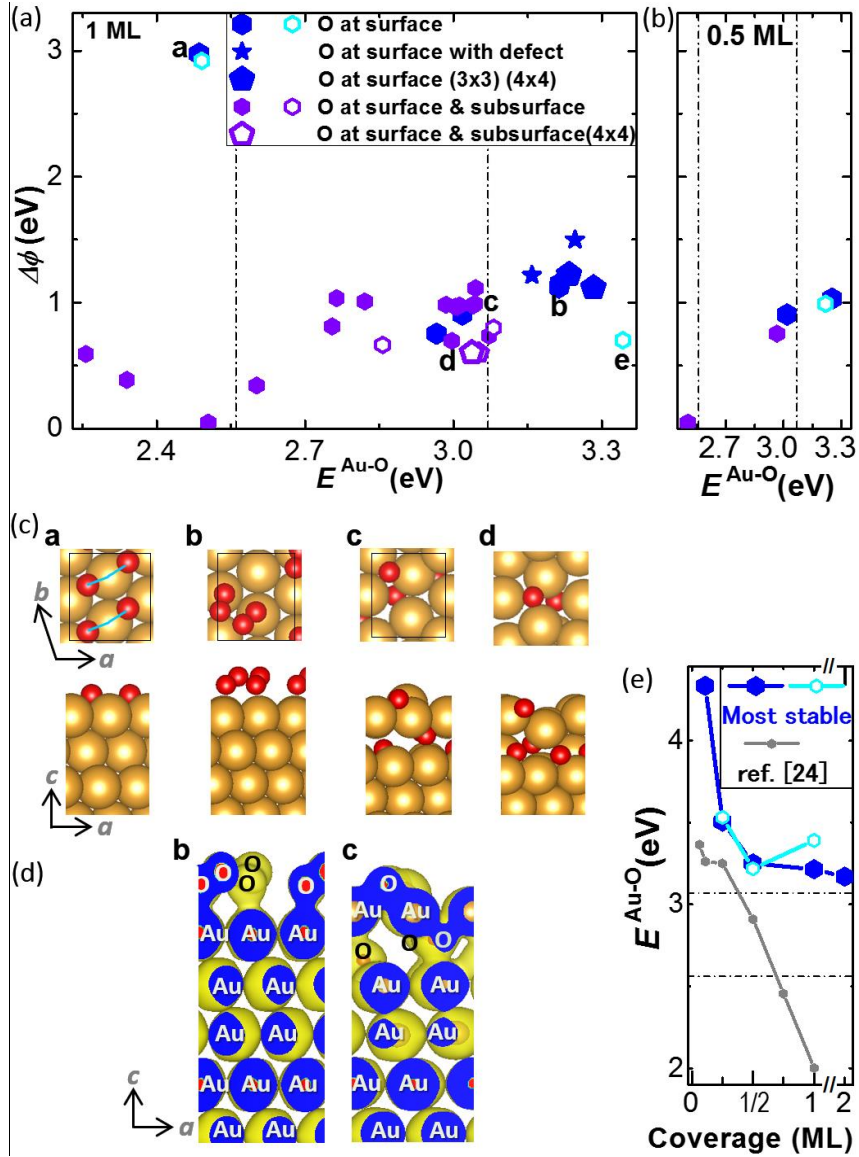

**Fig. 7.** $\phi$ vs. $E^{\text{Au-O}}$ at Au(111) surfaces at (a) 1 ML and (b) 0.5 ML O-coverages. These results used $(2 \times 2)$ supercells, except for pentagons that used $(3 \times 3)$ and $(4 \times 4)$ supercells. (c) Structures "a" – "d" correspond to the data "a" – "d" in (a). (d) Dark yellow and blue show iso-$e^-$-surfaces of "b" and "c" (0.045 $e^-/\text{Å}^3$) and their cross-sections, respectively. (e) The highest $E^{\text{Au-O}}$ at each coverage in this work, plotted with $E^{\text{Au-O}}$ of ref. [24].[a] In (a), (b), and (e), the experimental and theoretical $E^{\text{O-O}}$'s are shown by a pair of black dash-dot lines, blue and light blue symbols show the data for all O-atoms located at surfaces, and purple symbols show the data for some O-atoms located at subsurfaces, solid and open symbols correspond to the calculations using O_s and O$_{\text{ST}}$ potentials, respectively. The structure with defects in (a) is Fig. 2(h). The structures with $E^{\text{Au-O}} \approx 3.3$ eV in (b) are Figs. 2(f) and 2(g). [a] Reproduced with permission from Phys. Rev. B. **76**, 075327 (2007). Copyright 2007 American Physical Society.

Concerning accuracy in $\phi$ of a given structure, hybrid functional (e.g., HSEsol) estimates electronic properties more accurately [40,41,42,56] than GGA (e.g., PBEsol). Hence, using HSEsol, we reexamined $\Delta\phi$ at Au(111), Au(100), and Au(110) surfaces of supercells consisting of 1×1×10 Au unit-cells and O-atoms.

$\Delta\phi$ calculated using HSEsol was approximately <0.5 eV higher than $\Delta\phi$ calculated using PBEsol (Fig. 8). $\Delta\phi$ calculated using PBEsol with $O_{ST}$ PAW potential was slightly lower than the experimental $\Delta\phi$ (light blue hexagons in Fig. 7(a)). Furthermore, PBEsol calculations showed that the structures having $E^{Au-O}$ only 0.2 eV lower than $E^{Au-O}$ of the most stable structure exhibited $\Delta\phi$ lower than the experimental $\Delta\phi$ (Fig. 7(a)). Hence, HSEsol results were considered to agree with experiments [15], when $O_{ST}$ was used.

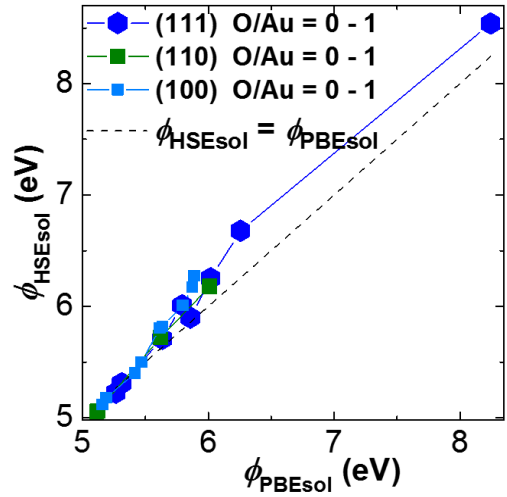

**Fig. 8.** Comparison of $\phi$ between PBEsol and HSEsol calculations at (111), (110), and (100) surfaces. Dashed lines show the curve of PBEsol $\phi$ = HSEsol $\phi$.

## D. Structures of O-covered Au surfaces

At low O-coverages, most stable structures of O-covered Au surfaces in this work agree with the previous DFT studies [17, 21, 24]. At high coverage, the most stable structures show new structures and evident rearrangements of Au-atoms. In particular, at Au(111) surfaces, we believe that the structures that hold O-atoms partly at subsurfaces mix into those that hold O-atoms only at surfaces and are most stable, which implies the loss of long-range orders. Details are in Secs. IIID1-IIID2.

### D1. Stable structures at low and high coverages

In the most stable structures of Au(111), Au(100), and Au(110) surfaces, O-atoms were at the surfaces. In these structures, at low coverages, O-atoms were dispersed and chemisorbed, as reported previously, but at high coverages, aggregated on occasion.

Specifically, at low O-atom coverages, we observed that Au-atoms were almost at the pristine Au positions of all the examined Au surfaces, e.g., Figs. 2(a), 2(d), and 2(e), "a" - "c" in Fig. 5(c), and "b" - "e" in Fig. 6(b). At high coverages, the structures corresponding to the minimum free energy showed low symmetry, which is evident in O-atoms in "c" of Fig. 5(c) (Au(100)), "a" of Fig. 6(c) (Au(110)), and "b" of Fig. 7(c) (Au(111)).

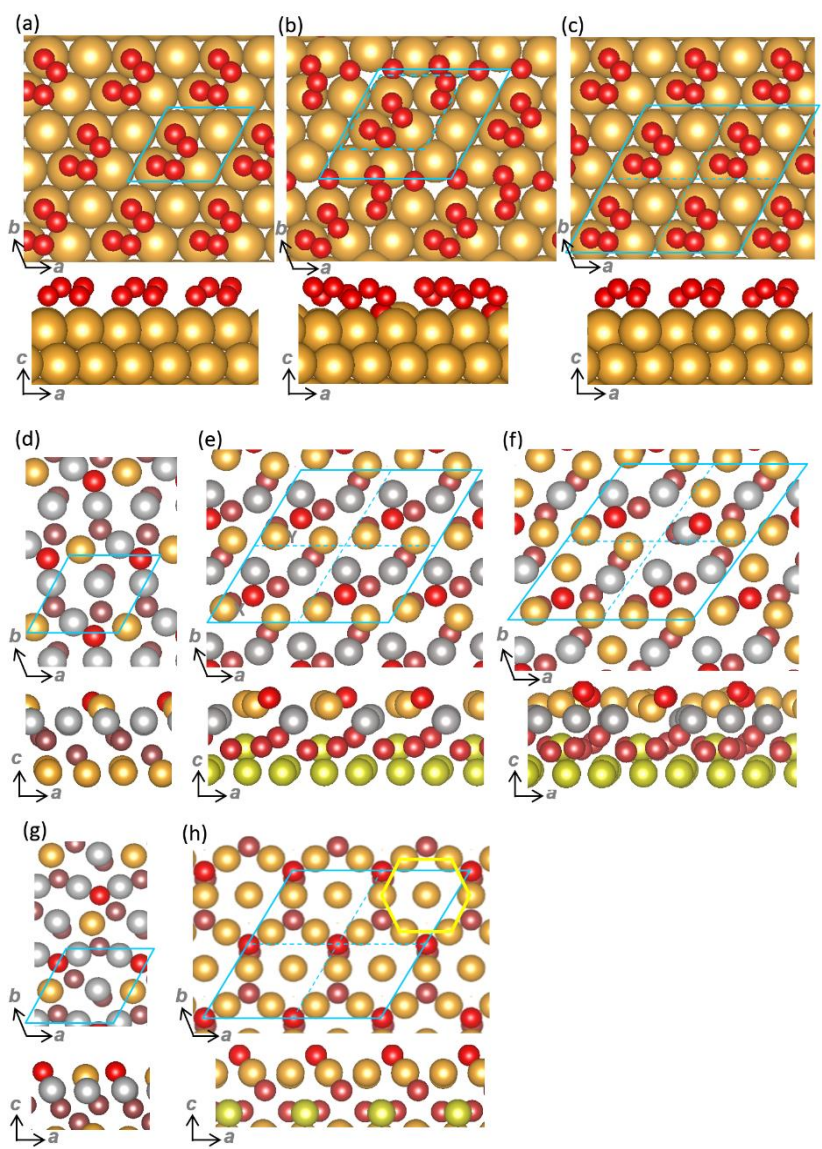

**Fig. 9.** Au(111) surface structures for minimum free energy, calculated using different supercell sizes shown by light blue parallelograms: O-atoms at surfaces (a)-(c) and at surfaces and sub-surfaces (d)-(h). (a), (d), and (g) 2×2. (b) 3×3. (c), (e), (f), and (h) 4×4. Dashed light blue lines show periodic units. The top views show only Au-atoms that existed at the surfaces of pristine Au. Dark yellow spheres in (e), (f), and (h) show Au-atoms that existed at the 2nd top layers of pristine Au. In (d)-(h), the diameters of the Au-atoms are reduced to visualize subsurface atoms. Otherwise, color and size conventions are the same as Figs. 2 and 5-7. The yellow hexagon in (h) marks a honeycomb structure.

In some cases, the 1$^{st}$ Au layer was split into two, as in Figs. 2(b) ((100)) and 2(f) ((111)) and "d" of Fig. 5(c) ((100)). Au-atoms were also displaced in-plane from the pristine positions as "a" in Fig. 6 (1 ML, (110)). In these cases, the surface Au-atoms reconstruct 2×1 and 2×2 surfaces at Au(100) and Au(110) surfaces, respectively. These reconstructions may be observable by low-energy electron diffraction (LEED). Additionally, "c" in Fig. 5 ((100)) and "b" in Fig. 7 ((111)) show molecule-like O arrangements; in all figures, the diameters of O-spheres are such that the O-atoms in O$_2$ are about to touch mutually (Fig. 2(i)). Except for Figs. 9(d)-9(h), the diameters of Au-spheres are such that the inner Au-atoms are about to touch mutually.

### D2. Molecule-like O-arrangements and disorder at Au(111)

When all O-atoms were at surfaces, the calculated final atom positions of Au(111) surfaces were not sensitive to the initial positions and converged to a few stable structures (Fig. 9(a)), which were mutually similar. These structures were the most stable, in which the O-atoms formed a pair of molecule-like aggregations that had the O-O distance slightly shorter than that of a free-standing O$_2$ molecule. These pairs were in not only the 2×2 supercell (Fig. 9(a)) but also 3×3 and 4×4 supercells (Figs. 9(b) and 9(c)); although these pairs were incompatible with 3×3 supercells, they also appeared in these supercells (Fig. 9(b)). Hence, we believe that these pairs of O-aggregations are ordering or the units of supercells.

The majority of data points in Fig. 7(a) is of the structures that contained a part of O-atoms at Au(111) subsurfaces. Some of these structures showed $E^{Au\text{-}O}$ that was only 0.2 eV lower than the highest $E^{Au\text{-}O}$ and comparable with $E^{O\text{-}O}$.

The structures of Au(111) surfaces that contained all the O-atoms at subsurfaces were the least stable and had $E^{Au\text{-}O}$ 1 eV lower than $E^{Au\text{-}O}$ of the most stable ones. The structures that contained 1/4 and 3/4 of O-atoms at subsurfaces (e.g., "c" and "d" in Fig. 7(c)) were more stable than those that contained 1/2 of O-atoms at subsurfaces. In these surfaces, the displacements of Au-atoms were evident, and the iso-$e^-$-density surfaces in "c" of Fig. 7(d) suggest strong covalent bonding between O and Au.

When O-atoms were at subsurfaces, the calculated final surface structures changed sensitively with initial surface ion positions, in contrast with the structures with all O-atoms at surfaces; the abundance of these structures that contained some O-atoms at subsurfaces (purple hexagons in Fig. 7(a)) was due to the slight difference in initial surface ion positions. Hence, final structures were various; Fig. 7(a) represents a fraction of these structures.

As a result, the arrangements of O- and Au-atoms at these surfaces changed, as 2×2 supercells changed to 4×4 supercells (Fig. 9(d) vs. Figs. 9(e) and 9(f), Fig. 9(g) vs. Fig. 9(h)). In Figs. 9(e), 4×2 superlattices emerged; e.g., the Au-atom at "X" moved to "Y". In Figs. 9(f), 4×4 superlattice emerged. In Fig. 9(h), superlattices remained as 2×2 but changed to honeycomb-like. Hence, the periodicity of the atom arrangement increased up to the size of the supercells, and we expect that this result also holds for the supercells having huge surfaces.

The types of structures that contain O-atoms in subsurfaces (Sec. IIIc2) are equal to the number of O-atoms and thus increase with the size of supercells. Moreover, in experiments, the arrival sites of O-atoms are random, which, in theories, corresponds to an infinite set of initial O-atom positions. This results in countless sets of theoretical final ion positions for the structures with O-atoms partly at subsurfaces. Hence, unlike the structures with all O-atoms at surfaces (i.e., most stable structures), the variations of the structures having O-atoms partly at subsurfaces rapidly increase with the size of the supercell surfaces. Hence, we believe that these variations are "disorders".

Consequently, near RT, these structures may appear experimentally owing to entropy despite slightly low $E^{\text{Au-O}}$'s. Hence, in the RT experiments of the O-covered Au(111) surfaces at high coverages, the structures having a part of O-atoms at subsurfaces are non-negligible and, eventually, mix with the most stable structures that can form a long-range order (Fig. 9(c)). Hence, the long-range ordering will not appear at RT. This inference matches the experimental conclusion "disordering of the substrate occurs at higher coverages" by Saliba et al. [15], in addition to the agreement on $\phi$.

## D3. Chemisorption vs. molecule-adsorption, wetting-nucleation, and $\phi$

The chemisorption of molecule-like O at the most stable Au(111) surfaces at 1 ML was ambiguous in the structure shown in "b" in Fig. 7(c). However, in the iso-$e^-$-density in "b" of Fig. 7(d), the contacts between O and Au were wider than in the inner part of Au. Further, the degree of the total $e^-$ transfer between the O-aggregate and Au surface was the same as that between a single O-atom and Au surface;

integrated $e^-$ density curves (Fig. 3(b)) showed that the $e^-$ occupation differed only 2.6% between the O-aggregate and the single O-atom. Here, "b" was calculated with O_s potential.

Moreover, the most stable structure calculated with $O_{ST}$ potential ("e" in Fig. 7(a)) also showed molecule-like O and the iso-$e^-$-density surfaces virtually the same as those of "b". Hence, the molecule-like O-aggregates ("b" and "e" of Fig. 7 and Fig. 9(a)) were considered chemically bound to Au. We remark that "e" exhibited the highest $E^{Au-O}$ and $\Delta\phi$ that best agreed with the experimental $\Delta\phi$ [15].

Thus, the side-views of structures (Figs. 2, 4, 5, 7, and 9), the binding energies, and the iso-$e^-$-density surfaces show that O-atoms at ≤1 ML were *chemisorbed* on Au(100), Au(110), and Au(111) surfaces. In contrast, the structures and iso-$e^-$-density surfaces in Fig. 10 indicated that O-atoms at 2 ML were rather molecular-adsorbed on these surfaces.

In the wetting theories of liquid on solid surfaces, wetting occurs by the balance between liquid-surface force and cohesive force. Similarly, in the theories of thin film growth that assumes high adatom mobility, adatoms aggregate when adatom–adatom interactions surpass those of the adatom with the surface (Volmer-Weber mode [60]), and adatoms are dispersed on surfaces, when adatom–adatom interactions are weaker than those of the adatom with the surface (Frank-van der Merwe mode [60]).

Au is inert, and the binding of Au with O is much lower than those of other metals with O. The molecule-like O-arrangements (O-aggregations) also suggested that the attractive force between O-atoms was non-negligible at high coverages, compared to the O-Au binding. This inference was supported by the emergence of molecule-like O-aggregations at a high ML (2 ML) on all of (100), (110), and (111) surfaces (Figs. 10(a)-10(f)). Further, the iso-$e^-$-density surfaces suggested the noticeable bonding between O-atoms and the reduced bonding between O and Au (Figs. 10(g)-10(i)).

The reduced bonding between O and Au implies the reduction of $e^-$-transfer between O and Au and, hence, the reduced electric dipole moments of O-Au per O-Au pair. The reduction explained the near-constancy of $\Delta\phi$ between 1 and 2 ML in Fig. 2(b), 5(a), and 6(b), which was consistent with experiments [15,23].

A longer distance between O and Au would enhance $\phi$ because the increase in $\phi$ by O-atoms is due to the effective dipole formed by O and Au atoms. However, in Fig. 7(a), $\phi$ of "a" was much higher than $\phi$ of "b", although the distance between O and Au in "a" was shorter than that in "b" (see also Fig. 3). This supported that the $e^-$ transfer between O and Au was important for $\phi$.

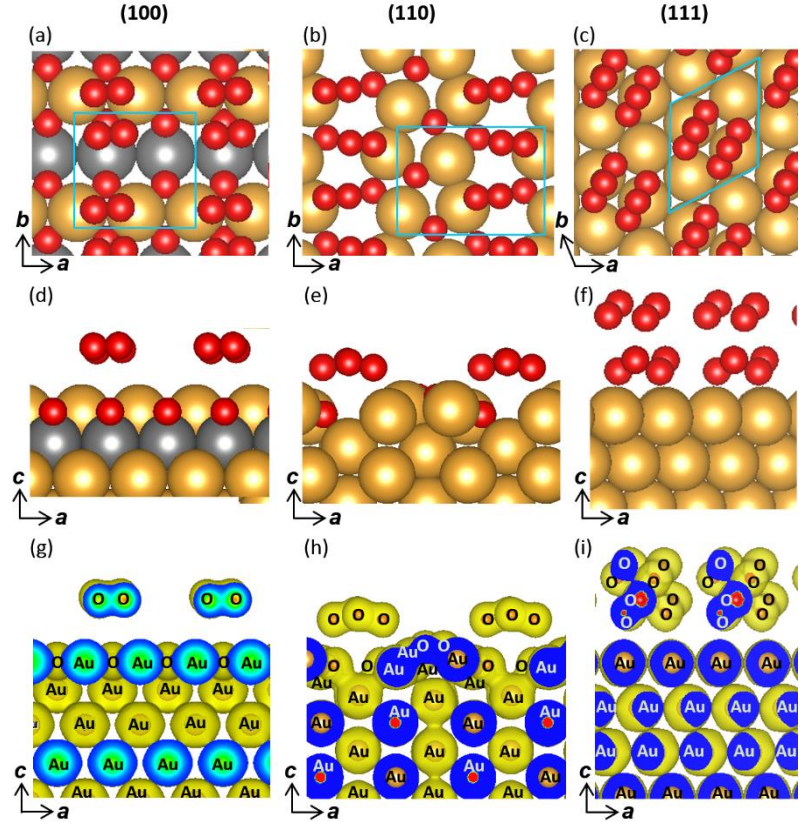

**Fig. 10.** Structures at 2 ML O-coverage: (a)-(c) top views and (d)-(f) side views. (g)-(i) Iso-$e^-$-density surfaces (0.045 $e^-/\text{Å}^3$). (a), (d), and (g): (100). (b), (e), and (h): (110). (c), (f), and (i): (111).

## E. Origin of differences between the present and previous DFT

At low coverages at Au(111) surface, $\Delta\phi$ reported in this work and ref. [24] matched the experiments [15], whereas at high coverages $\Delta\phi$ reported only by this work matched the experiments [15]. The origin of this difference is the O-atom aggregation at high coverage, which was unreported because refs. [17,21] studied O adsorptions at low coverages. We explain this in this section.

At low coverages, the sites of O adsorption were very near hole or bridge sites surrounded by Au-atoms; Au(100): Figs. 2(a) and 2(b) and "a" and "b" of Fig. 5, Au(110): Fig. 2(d), Au(111): Fig. 2(e), and "a" of Fig. 7. These results agree with those of Sun et al. [17] and Daigle and BelBruno [21].

Further, we also observed "linear O−Au−O" arrangements reported by Sun et al. [17], i.e., the structure of a freestanding $AuO_2$ molecule (Sec. IIIA3). They emerged mainly when Au-atoms were not much displaced from the pristine positions. These cases were when O-atoms were at high symmetry sites or low coverages.

In contrast to ref. [24], $E^{Au-O}$ of the most stable structures at each coverage in this work was almost unchanged and higher than $E^{O-O}$ obtained by experiments [48] and by DFT (Figs. 5(b), 6(b), and 7(e)). This $E^{Au-O}$ was consistent with the stable high-coverage surfaces in experiments [15]. $\Delta\phi$ of these surfaces in this work was <1.1 eV. These $E^{Au-O}$ and $\Delta\phi$ that were clearly different from those of ref. [24] result from the O-induced displacements of Au-atoms and the aggregation of O-atoms revealed by this work.

Indeed, when limited to high symmetry structures at Au(111) surfaces, $E^{Au-O}$ in this work also rapidly decreased with coverage and became small at high coverages, similar to ref. [24]. $\Delta\phi$ of these surfaces disagreed with experiments. These results of the high symmetry structures agreed with ref. [24]; $\Delta\phi$ and $E^{Au-O}$ of ref. [24] are in Figs. 1(a) and 7(e). In particular, $\Delta\phi$ and $E^{Au-O}$ of "a" in Fig. 7(a) for 1 ML matched those of ref. [24]. Hence, we believe that the results at ≥0.5 ML reported by Stampfl and a coworker [24] are of metastable structures.

In the search for stable Au(111) structures, ref. [24] "considered the hcp- and fcc-hollow sites" for "on-surface adsorption". Instead, to enforce the extensive searches for stable structures, we added the initial structures that contained O-atoms at unstable sites, using a strict convergence criterion on DFT. This approach is inefficient, but the geometry relaxation using all forces < 1 meV/Å occasionally showed that the final structures with all forces < 10 meV/Å used in ref. [24] were metastable and changed to different structures. Additionally, in contrast to ref. [24], we relaxed all the ions using thick supercells; we remark that upon O-chemisorption, even the Au-atoms in the 4th layer were displaced 0.01 Å (0.3%) in-plane and 0.03 Å (1.3%) out-of-plane.

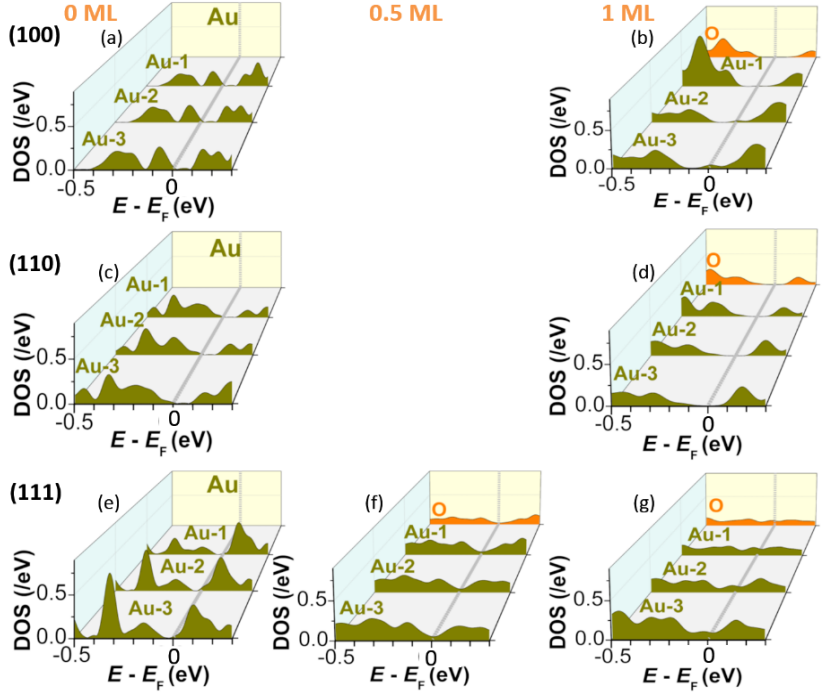

**Fig. 11.** LDOS of pristine and O-chemisorbed Au surfaces calculated using HSEsol. (a), (b) (100). (c), (d) (110). (e)-(g) (111). Au-1, Au-2, and Au-3 denote LDOSs of Au at the top, $2^{nd}$ top, and $3^{rd}$ top layers.

## F. DOS (*ab initio*)

Local DOS (LDOS) at surfaces is crucial for the image contrast of scanning tunneling microscopy [2] and for chemical reactivity. It is known that DFT, including GGA (e.g., PBEsol), noticeably underestimates band gaps, while HSEsol properly estimates band gaps [40-42]. Thus, we show LDOS calculated using HSEsol in Figs. 11. Owing to high computational costs, we show the HSEsol results only for symmetric initial O-atom positions.

Upon O-chemisorption, LDOSs at $E_F$ nearly vanished at the $1^{st}$ and $2^{nd}$ top layer at Au(100) and Au(110) surfaces and exhibited small gaps (Figs. 11(a)-11(d)), which was invisible at Au(111) surfaces (Figs. 11(e)-11(g)). These properties were unclear in PBEsol calculations.

The iso-$e^-$-density surfaces in Fig. 7(d) showed that $e^-$-density between atoms was higher near surfaces than insides, which suggested that LDOS near $E_F$ was greater at surfaces than within the bulk. This inference is because covalent bonding is considered primarily due to outer orbit $e^-$ and outer orbits are associated with high energy levels, i.e., near $E_F$. This enhancement of LDOS near surfaces is consistent with Fig. 12 that displays a marked correlation of LDOSs between $O_{2p}$ and $Au_{5d}$ at the top layers, as expected for $O_{2p}$-$Au_{5d}$ covalent bonding.

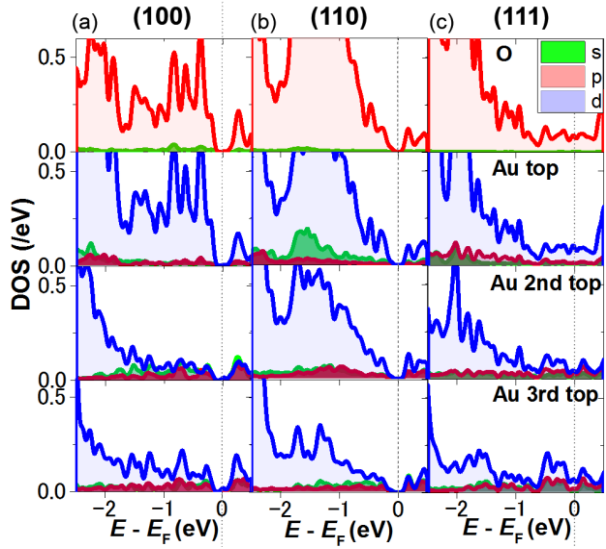

**Fig. 12.** s-, p-, and d-orbit-decomposed LDOS at 1 ML, corresponding to Figs. 11(b), 11(d), and 11(g). (a) (100). (b) (110). (c) (111). From top to bottom panels, the O-, the 1$^{st}$ Au-, the 2$^{nd}$ Au-, and the 3$^{rd}$ Au-layers.

## IV. SUMMARY

Theoretical and experimental $\phi$ can be quantitatively compared even in the absence of long-range orders. This invaluable feature facilitates the examination of the appropriateness of theoretical surface structures. Here, Au films are standard electrodes for metal oxide devices [61-63] and serve as a reference for $\phi$ in band diagrams [22,61-63], while oxides having Au electrodes are often $O_2$-treated.

In the reported [15,23] and present experiments, $\Delta\phi$ (the difference in the work function ($\phi$) between pristine and O-covered Au surfaces) was ≤1 eV (Figs. 1 and 6). Despite intensive studies of O-covered Au surfaces, theoretical studies of $\phi$ at these surfaces are scarce. An exception is the DFT on $\phi$ at Au(111) surfaces by Stampfl and a coworker [24] that shows $\Delta\phi \approx 3$ eV and $E^{Au\text{-}O} \approx 2$ eV at 1 ML ($E^{Au\text{-}O}$: formation energy of O-covered surface per O-atom), which is distinctly smaller than $E^{O\text{-}O}$ (binding energy of $O_2$ per O-atom). This $\Delta\phi$ [24] disagrees with the reported [15, 23] and present experiments (Figs. 1 and 6), and this $E^{Au\text{-}O}$ [24] is difficult to explain the experiments of O-chemisorption on Au in $O_2$ [5,6] that produces $\sqrt{3} \times \sqrt{3}$ superlattices suggesting a high or relatively high O-coverage.

Hence, this work studied oxygen-covered Au(100), Au(110), and Au(111) surfaces, regarding the experiments on $\phi$ as guides [15,23] for DFT and hybrid functional calculations. These calculations employed two kinds of atomic potentials for O, full-relaxations of all atom positions, and many variations of initial structures that included structures that contained O-atoms at subsurfaces. In particular, the inclusion of the initial structures containing unstable or metastable sites [17, 21] enforced the extensive searches of stable structures through a strict convergence criterion.

The most stable structures of Au(100), Au(110), and Au(111) surfaces showed $\Delta\phi < 1.1$ eV at $\leq 2$ ML, matching the experimental $\Delta\phi$ by Saliba et al. [15], Gottfried et al. [23], and the present experiments (Figs. 1(b) and 6(a)). These structures held all O-atoms at surfaces; O-atoms were chemisorbed at $\leq 1$ ML but considered close to be molecular-adsorbed at 2 ML.

In contrast with ref. [24], $E^{\text{Au-O}}$ of these structures was higher than the experimental and theoretical $E^{\text{O-O}}$ (Figs. 5-7) and can explain the O-chemisorption in $O_2$ [5,6]. When limited to low O-coverages, $\Delta\phi$, $E^{\text{Au-O}}$, and structures in this work agreed with the previous DFT [17,21,24].

At high coverage, we found that Au-atoms were noticeably displaced from pristine positions due to O-atoms and formed superlattices such as 2×1 (Figs. 5-7); e.g., the splitting or the corrugation of surface Au layers at 0.5 ML and 1 ML formed atomic-scale steps to maintain $E^{\text{Au-O}}$ (Figs. 2(b) and 2(f), "d" of Fig. 5, and Fig 6(c)).

At Au(111) surfaces at 1 ML, the most stable structures (having O-atoms at surfaces) exhibited an "ordering" consisting of molecule-like O (Figs. 9(a)-9(c)), which, by the analogy to wetting theories, suggested the non-negligible attractive interaction between O-atoms at high coverages. This interaction and the aforementioned O-induced displacements of Au-atom positions *altered the criteria for most stable surface structures at high O-coverages*. This transition in the criteria was considered the origin of the difference between the present and previous results [24].

Further, many quasi-stable structures containing O-atoms at subsurfaces were found at Au(111) surfaces at high coverages and showed a low $\phi$. These structures changed sensitively with the initial atom positions and the size of supercells. Hence, we believe these structures do not form long-range ordering. Because of the values of a relatively high $E^{\text{Au-O}}$ and the abundance of variations, we expect that these structures mix into the ordered structures, resulting in the loss of long-range order near RT. This expectation matches the experiments of Saliba et al. [15]. We may suggest that the consistency between the theoretical and experimental $\Delta\phi$ [15, 23] makes $\Delta\phi$ a reference property for O-coverage at Au surfaces.

DOS at each layer near surfaces calculated using hybrid functional showed small gaps at O-chemisorbed Au(100) and Au(110) surfaces, which were absent at Au(111) surfaces.

**Acknowledgements**

This work was supported partly by Murata Science and Education Foundation. The discussions with P. Blöchel of TU Clausthal is greatly acknowledged.

**Declaration of competing interest**      The author declares no competing interest.

**Data Availability**

The data in my manuscript can be obtained from the corresponding author upon reasonable request.